\documentclass[10pt,conference]{IEEEtran}%

\usepackage[letterpaper, left=1in, right=1in, bottom=1in, top=0.75in]{geometry}
\usepackage{array}
\usepackage[cmex10]{amsmath}

\usepackage{amssymb}   
\usepackage{amsxtra}
\usepackage{amscd}
\usepackage{amsthm}
\usepackage{amsxtra}
\usepackage{amscd}
\usepackage{amsthm}
\usepackage{tikz}
\usetikzlibrary{matrix}
\usepackage{mathtools}
\usepackage{graphicx}   
\usepackage{wrapfig}
\usepackage{layouts}
\usepackage{array}  
\usepackage{multirow}  
\usepackage{tabularx}
\usepackage{arydshln}
\usepackage{mathtools}
\usepackage{xparse}

\NewDocumentCommand{\overarrow}{O{=} O{\uparrow} m}{%
	\overset{\makebox[0pt]{\begin{tabular}{@{}c@{}}#3\\[0pt]\ensuremath{#2}\end{tabular}}}{#1}
}
\NewDocumentCommand{\underarrow}{O{=} O{\downarrow} m}{%
	\underset{\makebox[0pt]{\begin{tabular}{@{}c@{}}\ensuremath{#2}\\[0pt]#3\end{tabular}}}{#1}
}

\usepackage{verbatim}
\usepackage{stfloats}
\usepackage{blindtext}
\usepackage{abraces}

\usepackage{color}

\usepackage{float}
\usepackage{fixltx2e}
\usepackage{xparse}
\usepackage{cite}
\renewcommand{\thefootnote}{\arabic{footnote}}
\usepackage[keeplastbox]{flushend}

\usepackage{color}
\usetikzlibrary{decorations.pathreplacing,
	matrix,
	positioning,
}

\usepackage[bottom]{footmisc}
\usepackage{tikz}

\usetikzlibrary{decorations.pathreplacing}
\newcommand{\vast}{\bBigg@{4}}
\newcommand{\Vast}{\bBigg@{5}}

\let\svthefootnote\thefootnote

\usepackage{comment}

\begin{document}
	\emergencystretch 3em
	\title{{SimMBM Channel Simulator for Media-Based Modulation Systems}}

	\author{\IEEEauthorblockN{Zehra Yigit\textsuperscript{$\dagger$}, Ertugrul Basar\textsuperscript{$*$}, and Ibrahim Altunbas\textsuperscript{$\dagger$}
			\IEEEauthorblockA{\textsuperscript{$\dagger$}Istanbul Technical University, Faculty of Electrical and Electronics Engineering,  Maslak, 34469, Istanbul, Turkey}}
		\IEEEauthorblockA{\textsuperscript{$*$}CoreLab, Department of Electrical and Electronics Engineering, Ko\c{c} University, Sariyer, 34450, Istanbul, Turkey. \\
			E-mail: yigitz@itu.edu.tr, ebasar@ku.edu.tr,  ibraltunbas@itu.edu.tr}
		
	}

	
	%


	\maketitle
	
	\begin{abstract}
		Media-based modulation (MBM), exploiting rich scattering properties  of transmission environments via different radiation patterns of a single reconfigurable antenna (RA), has brought new insights into  future communication systems. In this study, considering this innovative transmission principle, we introduce the realistic, two-dimensional (2D), and open-source \textit{SimMBM} channel simulator to support  various applications of   MBM  systems at sub-$6$ GHz frequency band in different environments.

	\end{abstract}
	

	%
	\IEEEpeerreviewmaketitle
	\let\thefootnote\relax\footnote{This work was supported  by TUBITAK under grant no 117E869.
		MATLAB script of SimMBM Channel Simulator is
		available at https://corelab.ku.edu.tr/tools}
	\addtocounter{footnote}{-1}\let\thefootnote\svthefootnote
	
	\section{Introduction}
	The fifth generation (5G) technology envisions  enabling numerous new applications    with diverse characteristics including low-latency, high reliability, high data rate, energy efficiency and security under three core use-cases: ultra reliable low-latency communications (uRLLC), enhanced mobile  broadband (eMBB) and massive machine type communications (mMTC) \cite{series2017guidelines}. While mMTC aims to provide connectivity of the massive number of devices that intermittently  send  or receive small amount of data, uRLLC applications are anticipated to deal with  the stringent requirements on ultra reliability and ultra low-latency for mission critical transmissions. On the other hand, eMBB targets extended coverage, extreme data rates and huge amount of data transfer.
	
	Multiple-input multiple-output (MIMO) systems play a pivotal role  to overcome the high data rates requirements of 5G and beyond technologies   \cite{series2017guidelines}.  Although, millimeter wave (mmWave) and massive MIMO  (mMIMO) systems are considered as promising transmission techniques,  their inevitably high energy consumption  leads researchers to consider more energy efficient solutions \cite{basar2019wireless}. As these studies continue to expand, 	index modulation (IM) techniques has emerged as a  challenging  candidate for the future communication technologies, which  achieve  high data rates by  exploiting indices of main pillars of a basic communication system in a clever manner to convey extra information \cite{basar2019wireless}.    

	Media based modulation (MBM), which is one of the prominent IM techniques and uses rich scattering characteristics  of  wireless environments via exploiting the signatures of different radiation  patterns of a reconfigurable antenna (RA), adds a wholly new aspect to future  wireless transmission technologies. 


	This creative transmission technique has  attracted a growing  interest in the literature. In \cite{khandani2014media},  the authors emphasize  significant gains of the MBM system using RF mirrors for   practical channel realizations compared to classical single-input multiple-output (SIMO) and MIMO systems. In the follow-up studies, different space-time block coding (STBC) techniques are integrated with the  MBM transmission scheme to achieve numerous orders of  transmit diversity gains \cite{basar2017space, yigit2019space}. Also,  multi-user  and massive MIMO  implementations of the MBM system are reported in  \cite{shamasundar2018media}.   For further discussion, we refer the interested reader  to \cite{basar2019media} and the references therein.  

	\begin{table*}

		\centering
		\caption{Channel Parameters for sub-$6$ GHz Spectrum }
		
		\begin{tabular}{|c|c|c|c|c|c|c|c|}
			\hline
			\multicolumn{2}{|c|}{\multirow{2}{*}{}} & \multicolumn{2}{c|}{\bf{InH}} & \multicolumn{2}{c|}{\bf{UMi}} & \multicolumn{2}{c|}{\bf{UMa}} \\ \cline{3-8} 
			\multicolumn{2}{|c|}{}                  & \bf{\text{LOS}}          &  \bf{NLOS}             &    \bf{\text{LOS}}         &    \bf{NLOS}           &       \bf{\text{LOS}}         &     \bf{NLOS}           \\ \hline
			\multirow{2}{*}{{Delay spread  (DS)} $\log_{10}({s}$) }       &    $\mu$       &  $-7.70 $    &  $-7.41 $    &       $-7.19 $        &     $-6.89 $      &  $ -7.03$   & $-6.44$            \\ \cline{2-8} &
			$\sigma$&    $0.18$       &       $0.14$     &   $0.40$         &    $0.54$                &       $0.66$     &     $0.39$        \\ \hline
			\multirow{2}{*}{AoD spread  (ASD) $\log_{10}({^\circ}$) }       &    $\mu$       &  $1.60 $    &  $1.62 $    &       $1.20 $        &     $1.41 $      &  $ 1.15$   & $1.41 $            \\ \cline{2-8} &
			$\sigma$&    $0.18$       &       $0.25$     &   $0.43$         &    $0.17$                &       $0.28$     &     $0.28$        \\ \hline
			\multirow{2}{*}{AoA spread  (ASA) $\log_{10}({^\circ}$) }       &    $\mu$       &  $1.62 $    &  $1.77 $    &       $1.75 $        &     $1.84 $      &  $ 1.81$   & $1.87 $            \\ \cline{2-8} &
			$\sigma$&    $0.22$       &       $0.16$     &   $0.19$         &    $0.15$                &       $0.20$     &     $0.11$        \\ \hline
			\multicolumn{2}{|c|}{AoA/AoD distribution }             &
			\multicolumn{2}{c|}{Laplacian } & \multicolumn{2}{c|}{Wrapped     Gaussian} & \multicolumn{2}{c|}{Wrapped     Gaussian} \\ \hline
			\multicolumn{2}{|c|}{Number of clusters }             &    { $15$  } & {  $19$}  & { $12$  }       &   {   $19$}   &  {  $14$}      &      {$12$}     \\ \hline
			\multicolumn{2}{|c|}{Number of rays per cluster }             &    {$20$ }  & {  $20$}  & { $20$ }        &  { $19$ }  &     {$20$}      & {     $20$ }    \\ \hline
			\multicolumn{2}{|c|}{Cluster ASD }            &    $5$       &     $5$      &    $3$       &       $10$    &      $5$     &  $2$         \\ \hline
			\multicolumn{2}{|c|}{Cluster ASA }             &    $8$       &     $11$      &    $17$       &       $22$    &      $11$     &  $15$   \\\hline      
		\end{tabular}
		\label{t2}
	\end{table*}

	Reconfigurable intelligent surface (RIS) technology is also considered as a potential application  to deal with  energy efficient and cost-effective network requirements of  5G and beyond systems \cite{basar2020indoor}. An RIS adapts the propagation environment of  a communication system in a constructive manner through controllable passive electromagnetic  materials that only induce  phase shifts without any power supply and complicated signal processors \cite{basar2019transmission}.  These low-cost  and practical use of RIS has led proliferation of RIS-aided communication systems under diverse research areas \cite{basar2019wireless}. Indeed,  recent studies present  physical channel models for RIS-aided  mmWave  communication systems \cite{basar2020recon, basar2020indoor, he2020large},  while  \cite{basar2020recon} and  \cite{basar2020indoor} introduce an open-source SimRIS channel simulators to   model propagation characteristics of mmWave channels in presence of RISs.     

	In this study, considering International Mobile Telecommunications-Advanced (IMT-Advanced) recommendations  \cite{series2009guidelines}, the \textit{SimMBM} channel simulator is released for simulating  realistic and two-dimensional (2D)  MBM transmission channels  that are pertinent to  sub-$6$ GHz \cite{series2009guidelines} applications   of  specific indoor hotspot (InH), urban micro (UMi) and urban macro (UMa)  network configurations. Unlike the classical MBM studies \cite{basar2019media, basar2017space, yigit2019space,  shamasundar2018media} that consider RF mirrors around transmit antenna(s) to create questionably independent channel state realizations, in \textit{SimMBM}, we  utilize  a real RA  {with four radiation patterns \cite{jin2018simple} that inherently capable of  generating channel variations, while a generalization to the case of multiple radiation patterns is straightforward.} This open-source  channel simulator  offers different environment  types (InH, UMi and UMa), and   allows  locations variations. 




	The remainder of the paper is organized as follows. We present the overall structure of the channel simulator, including the MBM systems and  IMT-Advanced channel modeling concepts  in Section II. In Section III, the channel correlation and capacity measurements of the MBM systems are provided. Our numerical results are presented in Section IV and conclusions are given in Section V. 
	
\enlargethispage{-1\baselineskip}
\begin{table}[!b]
	\vspace{-0.6cm}
	\centering
	\caption{LOS Probability  \cite{series2009guidelines} }
	\begin{tabular}{l |l }
		\hline	
		&  LOS Probability \\ \hline	\hline
		InH	&  $p(d)=\begin{cases} 1& d\leq18\\ \exp(-(d-18)/27)& 18<d<27\\
			0.5& d\geq37 \end{cases}$\\ \hline
		UMi	& $p(d)=\min(18/d,1)(1-\exp(-d/36))+\exp(-d/36) $\\ \hline
		UMa	& $p(d)=\min(18/d,1)(1-\exp(-d/63))+\exp(-d/63) $ \\ \hline 
	\end{tabular}
	\label{t1}
\end{table}		

	\section{SIMMBM Channel Simulator}
	In this section, the \textit{SimMBM} channel simulator, which generates  the   MBM  transmission channels characterized by  the  IMT-Advanced standards at sub-$6$ GHz frequency band, is introduced. 
	
	\subsection{IMT-Advanced Channel Modeling}
\begin{figure}[t]
		\vspace*{-0.5cm}	
		\noindent
		\makebox[\columnwidth][c]{\includegraphics[ width=0.95\columnwidth]{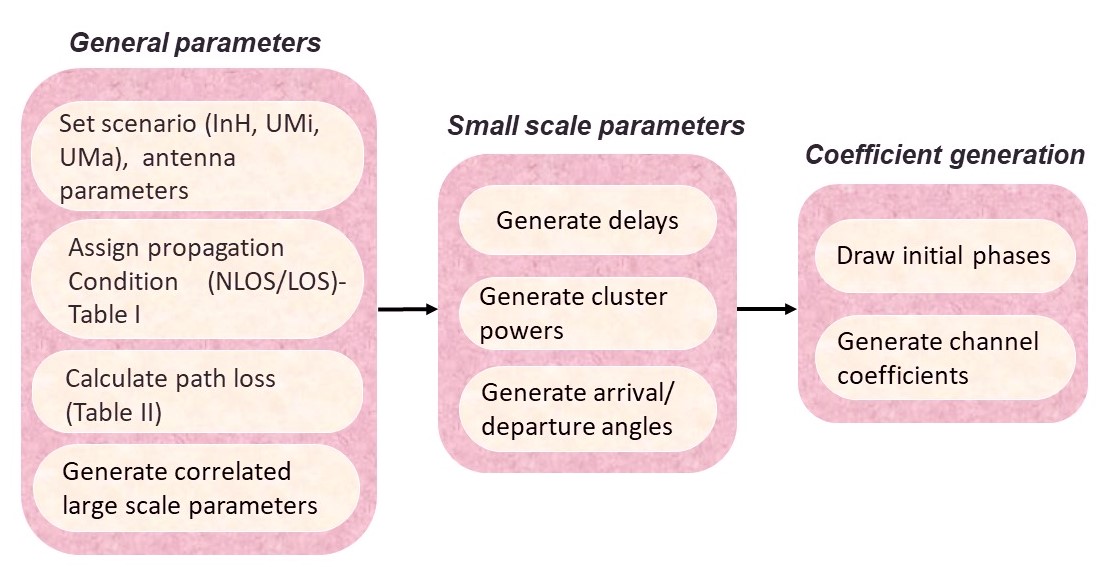}}
		\vspace{-0.3cm}
		\caption{IMT-Advanced channel coefficient generation procedure \cite{series2009guidelines}}
		\label{figx}
			\vspace{-0.7cm}
	\end{figure}

	The IMT-Advanced systems proving extensive  transmission protocols  for mobile communications and broadband connectivity are  growingly deployed worldwide in today's wireless communication systems \cite{series2017guidelines}.
	In this study, the proposed \textit{SimMBM} channel simulator, which produces  MBM channels in various indoor and outdoor environments, follows IMT-Advanced channel modeling procedure  illustrated in Fig. \ref{figx}  \cite{series2009guidelines}. 

%
	In this study, first, { we consider a generic single-input single-output (SISO) system whose propagation environment is characterized by   IMT-Advanced  channel models that consist of   multiple clusters, each composing a number of  sub-rays with similar characteristics referred to as \textit{scatterers}.} Each cluster is characterized by  \textit{large scale parameters}  which are statistical parameters such as delay spread (DS) and angular spread (AS) that follow the distributions given in Table \ref{t2} \cite{series2009guidelines}.  Then, these parameters are employed to assign delay $\tau_{c,s}$, power $P_{c,s}$,  departure $\phi_{c,s}$ and arrival  $\varphi_{c,s}$ angles to each path, which are often referred to as \textit{small scale parameters}.
	
\enlargethispage{-0.5\baselineskip}
	\begin{table}[!b]
	\vspace*{-0.8cm}
	\caption{Path loss Parameters \cite{series2009guidelines} }
	\begin{center}
		\begin{tabular}{l|l|c|c|c|c}
			\hline
			\multicolumn{2}{c|}{Scenario}                      & $K$ &  $L$& $M$ & $\sigma_{SF}$  \\ \hline\hline
			\multicolumn{1}{l|}{\multirow{2}{*}{InH}} & LOS & $16.9$  & $32.8$  & $20$ & $3$  \\ \cline{2-6} 
			\multicolumn{1}{l|}{}                  & NLOS  & $43.3$ & $11.5$  & $20$  & $4$  \\ \hline
			\multicolumn{1}{l|}{\multirow{2}{*}{UMi}}  & LOS & $22$  & $28$  & $20$ & $3$   \\ \cline{2-6} 
			\multicolumn{1}{l|}{}                  &  NLOS & $36.7$  & $22.7$  & $26$ & $4$  \\ \hline
			\multicolumn{1}{l|}{UMa}                  &  LOS & $22$  & $28$  & $20$ & $4$  \\ \hline
		\end{tabular}
	\end{center}
	\label{t3}
\end{table}

	Accordingly, after  specifying a scenario among the available network configurations  InH, UMi and UMa, distance-dependent {  line of sight (LOS) probability $p(d)$,    for  $d$ being the 2D distance between the transmitter  (T) and the receiver (R), is determined  as given in Table \ref{t1}} \cite{series2009guidelines}. 	{Please note that throughout this paper, the terms "T" and "R" will be used to refer transmitter and the receiver, respectively. }
	
	After  determining  whether a LOS propagation link exists between the T and the R,  the attenuation caused by path loss and shadowing fading (SF), in decibel (dB), is calculated,  using the  path loss parameters given  in Table \ref{t3} \cite{series2009guidelines}, as follows
	\begin{equation}
		P_L(d)\hspace{0.05cm}\text{[{dB}]}=-K\log_{10}(d)-L-M\log_{10}(f_c)-\sigma_{SF}
		\label{pl}
	\end{equation}
		where $f_c$ is operating frequency at  the sub-$6$ GHz frequency band.
		
	Assume $C$ and $S$  respectively represent the number of total clusters and number of scatterers within  each cluster, and $(c,s)$  denotes the $s$th  scatterer of the $c$th cluster. Therefore,  considering  IMT-Advanced channel parameters in Table \ref{t2}  \cite{series2009guidelines},  the geometric channel  between this generic SISO system is constructed  as   \cite{series2009guidelines}
	\begin{equation}
		h=\sum_{c=1}^{C}\sum_{s=1}^{S}\alpha_{c,s}\sqrt{G_t(\phi_{c,s})G_r(\varphi_{c,s})P_L(d)}e^{j\eta}+h_{\text{LOS}}
		\label{siso} 	
	\end{equation}
	where $h_{\text{LOS}}$ is the LOS component, $\eta$ is the initial phase uniformly distributed between $[0,2\pi]$, i.e., $\eta\sim\mathcal{U}(0,2\pi)$, while  ${G_t(\phi_{c,s})}$ and ${G_r(\phi_{c,s})}$ are the transmit and receive  antenna gains in the direction of the $(c,s)$th path, respectively.  In (\ref{siso}),  $\alpha_{c,s}\sim\mathcal{CN}(0,P_{c,s})$ is assumed to be independent identically distributed (i.i.d) complex Gaussian random variable with zero mean and $P_{c,s}$ variance  for $P_{c,s}$ being the average power of the $(c,s)$ path and $\sum\limits_{c,s} P_{c,s}=1$, while $h_{\text{LOS}}$  is determined as
	\begin{equation}
		h_{\text{LOS}}=I_{\text{LOS}}(d)\sqrt{G_t(\phi_{\text{LOS}})G_r(\varphi_{\text{LOS}})P_L(d)}e^{j\eta}
	\end{equation}	
	where $I_{\text{LOS}}(d)$ is Bernoulli random variable with $p(d)$ probability, indicating existence of a LOS link  between the T and R.   $G_t(\phi_{\text{LOS}})$ and $G_r(\varphi_{\text{LOS}})$ are the transmit and the received antenna gain in LOS direction, respectively.

	
	
	\begin{figure}[t]
		\centering
		\includegraphics[width=1\linewidth]{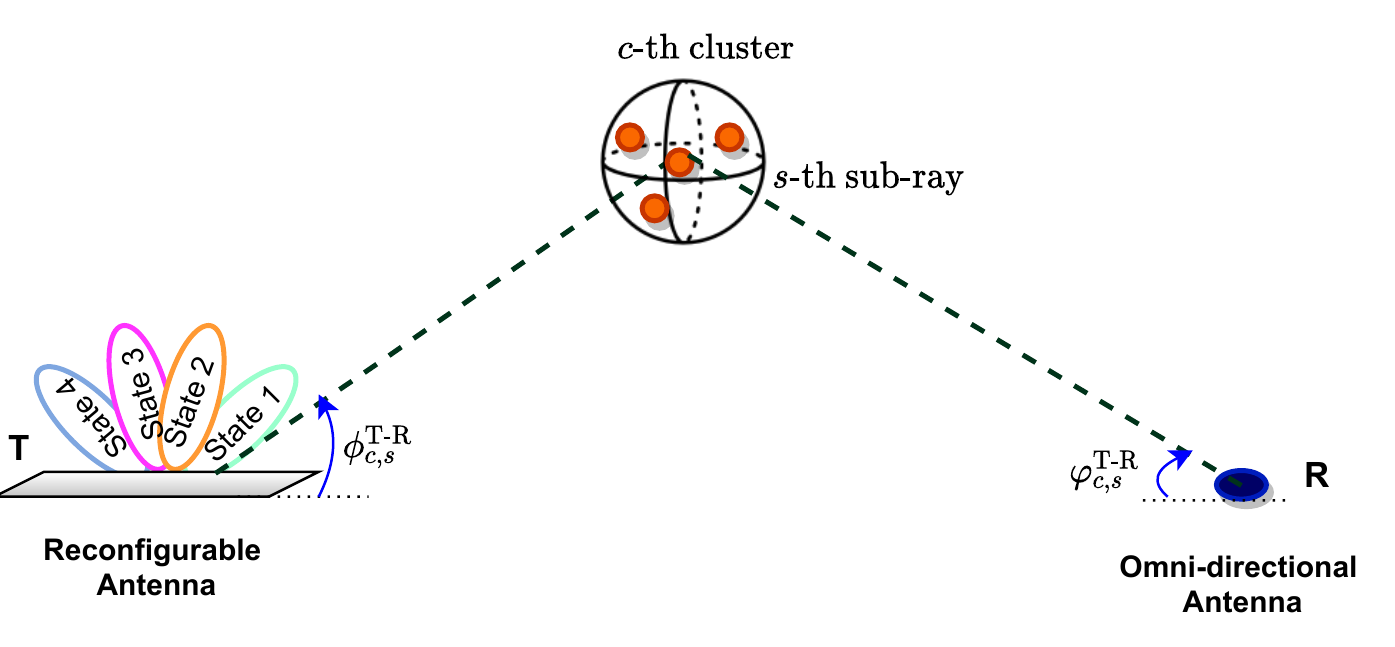}
		\vspace*{-0.6cm}
		\caption{SISO-MBM  schematic}
		\label{fig1}
		\vspace*{-0.5cm}
	\end{figure}	
\vspace{-0.2cm}
	\subsection{RA-based MBM   Channels  }
	In this study,  considering  IMT-Advanced standards for sub-$6$ GHz  applications described above \cite{series2009guidelines}, we consider propagation environment of a SISO system whose schematic diagram is given in Fig. \ref{fig1}.  It is assumed that a single RA with four different directional radiation patterns operating at $2.45$ GHz frequency \cite{jin2018simple} is employed at T, while  R is equipped with an omni-directional antenna that collects signals with unit gain in all directions. 
	
\begin{figure}[t]
		\centering
		\includegraphics[width=0.9\linewidth]{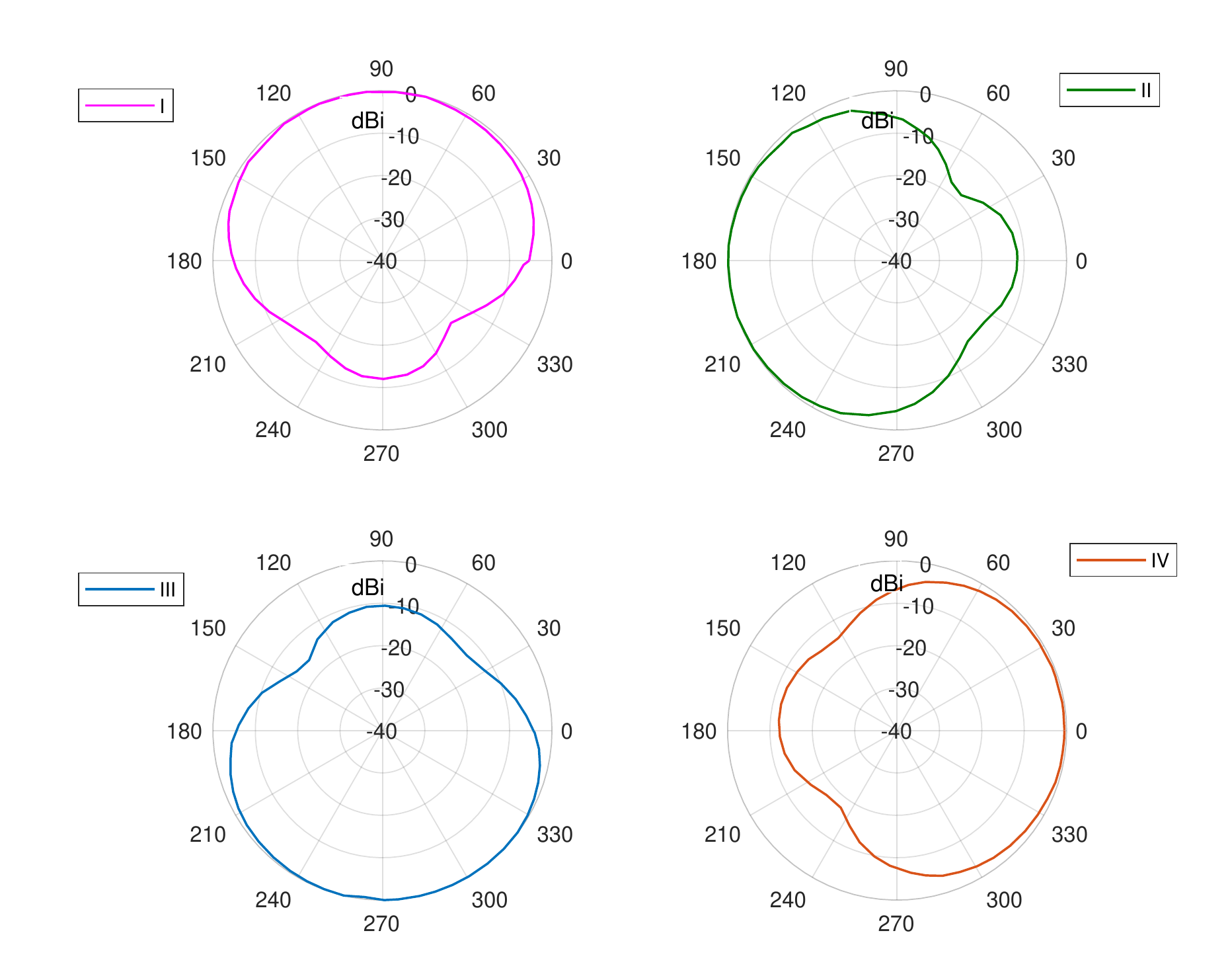}
			\vspace*{-0.3cm}
		\caption{ Radiation patterns of the four-state RA \cite{jin2018simple} in azimuth direction}
		\label{figure2}
			\vspace*{-0.7cm}
	\end{figure}

	The  considered RA exploits   four arc dipoles  as  radiating elements and manipulates  the radiation direction by ON/OFF status of PIN  diodes \cite{jin2018simple}, which  create  four directional radiation patterns presented in Fig. \ref{figure2}. It is worth-noting that the radiation patterns of this RA are  extracted  from 2D polar  plots  given in \cite{jin2018simple} by using a software program \cite{wpd}. 	Moreover,   this RA with pattern diversity can steer $360^\circ$ in azimuth plane  and allows us to perform the classical MBM concept \cite{naresh2016media,basar2019media}  between T and R. Therefore,   considering  the above azimuth beam steering  
	RA  \cite{jin2018simple}, an RA-based SISO-MBM system is constructed. 

	In this system, considering classical MBM transmission principle, the incoming $m=\log_2(4)=2$ information bits determine  one out of  four radiation  patterns of the RA \cite{jin2018simple} to be activated  in each transmission interval by switching ON/OFF status of the corresponding PIN diode. Therefore, considering the geometry-based IMT-Advanced channels reported in preceding subsection,  the transmission channel of the SISO-MBM system is constructed as  
	\begin{equation}
		h_p=\sum_{c=1}^{C}\sum_{s=1}^{S}\alpha_{c,s}\sqrt{G_p(\phi_{c,s})P_L(d)}e^{j\eta}+h_{\text{LOS}}
		\label{sisombm}
	\end{equation}
	where $G_p(\phi_{c,s})$ is the antenna gain of the $p$th state in direction of $\phi_{c,s}$ angle for  $p$ denoting the index of the active radiation pattern. 	

	{Furthermore}, to enhance the spectral efficiency of the  SISO-MBM scheme, it  is generalized for a MIMO  configuration  with uniform linear array (ULA) deployment of $T_x$ RA antennas at T  and $R_x$ omni-directional antennas at R. 
	 Further, in this MIMO-MBM system, by assigning $\log_2(4)=2$ bits to each RA,  all available transmit RAs are used for transmission through their active radiation patterns which are determined via  the  incoming $m=T_x\log_2(4)=2T_x$  bits. Therefore, the channel coefficient between the $i$th  transmit RA and $j$th received antenna is constructed as follows
	\begin{align}
		&\hspace{0cm}{f}_{i,j}=	\sum_{c=1}^{C}\sum_{s=1}^{S}\alpha_{c,s}\sqrt{{P_{L}(d_{\text{T-R}})}}\sqrt{G_{p}^i(\phi_{c,s}^{\text{T-R}})}
		\nonumber \\ &\hspace{1.5cm}\times e^{jkd_a(i-1)\sin(\phi_{c,s}^{\text{T-R}})} e^{jkd_a(j-1)\sin(\varphi_{c,s}^{\text{T-R}})}.
		\label{mimo1}
	\end{align} 
	In (\ref{mimo1}), for $d_{\text{T-R}}$ being distance between the T and R, $G_{p}^i(\phi_{c,s}^{\text{T-R}})$ is the  gain of the $p$th pattern corresponding to $i$th transmit RA,  $k=2\pi/\lambda$, and  $d_a$ is the  distance between adjacent antennas, which is assumed to be $d_a=\lambda/2$, {and   $\lambda$ is the wavelength.} Therefore, the overall channel matrix between T and R, $ \mathbf{F}\in\mathbb{C}^{R_x\times T_x}$, becomes
	\begin{align}
		\mathbf{F}=
		\sum_{c=1}^{C}\sum_{s=1}^{S}\alpha_{c,s}\sqrt{{P_{L}(d_{\text{T-R}})}}\mathbf{a}^{\mathrm{H}}(\phi_{c,s}^{\text{T-R}})\mathbf{a}(\varphi_{c,s}^{\text{T-R}})+{\mathbf{F}}_{\text{LOS}}
		\label{mimo}
	\end{align} 
	where ${\mathbf{F}}_{\text{LOS}}$ is the LOS component, $\mathbf{a}(\phi_{c,s}^{\text{T-R}})$ and $\mathbf{a}(\varphi_{c,s}^{\text{T-R}})$ denote the transmit and received array response vectors at the corresponding departure $\phi_{c,s}^{\text{T-R}}$ and arrival $\varphi_{c,s}^{\text{T-R}}$ angles, respectively. In (\ref{mimo}), the ULA-based RA array response  vector $\mathbf{a}(\phi_{c,s})\in\mathbb{C}^{T_x\times 1}$ { (including gains)} is
	\begin{align}
		&	\mathbf{a}(\phi_{c,s}^{\text{T-R}})=\Big[\begin{matrix}
			\sqrt{G_{p}^1(\phi_{c,s}^{\text{T-R}})}&\sqrt{G_{p}^{2}(\phi_{c,s}^{\text{T-R}})}e^{jkd_a\sin(\phi_{c,s}^{\text{T-R}})}&\cdots\end{matrix}\nonumber\\
		&\hspace{2.2cm}\begin{matrix}
			\sqrt{G_{p}^{T_x}(\phi_{c,s}^{\text{T-R}})}e^{jkd_a(T_x-1)\sin(\phi_{c,s}^{\text{T-R}})}
			\label{1mimo}
		\end{matrix}\Big]^T
	\end{align}
	while  ULA-based omni-directional receiver array vector $\mathbf{a}(\varphi_{c,s}^{\text{T-R}})\in\mathbb{C}^{R_x\times 1}$ is 
	\begin{equation}
		\mathbf{a}(\varphi_{c,s}^{\text{T-R}})=\begin{bmatrix}
			1&e^{jkd_a\sin(\varphi_{c,s}^{\text{T-R}})}&\cdots&e^{jkd_a(R_x-1)\sin(\varphi_{c,s}^{\text{T-R}})}
		\end{bmatrix}^T.
		\label{2mimo}
	\end{equation}
\vspace*{-0.8cm}
	\subsection{RIS-aided MBM Systems}  
\begin{figure}[t]
		\centering
		\includegraphics[width=1\linewidth, height=10cm,keepaspectratio]{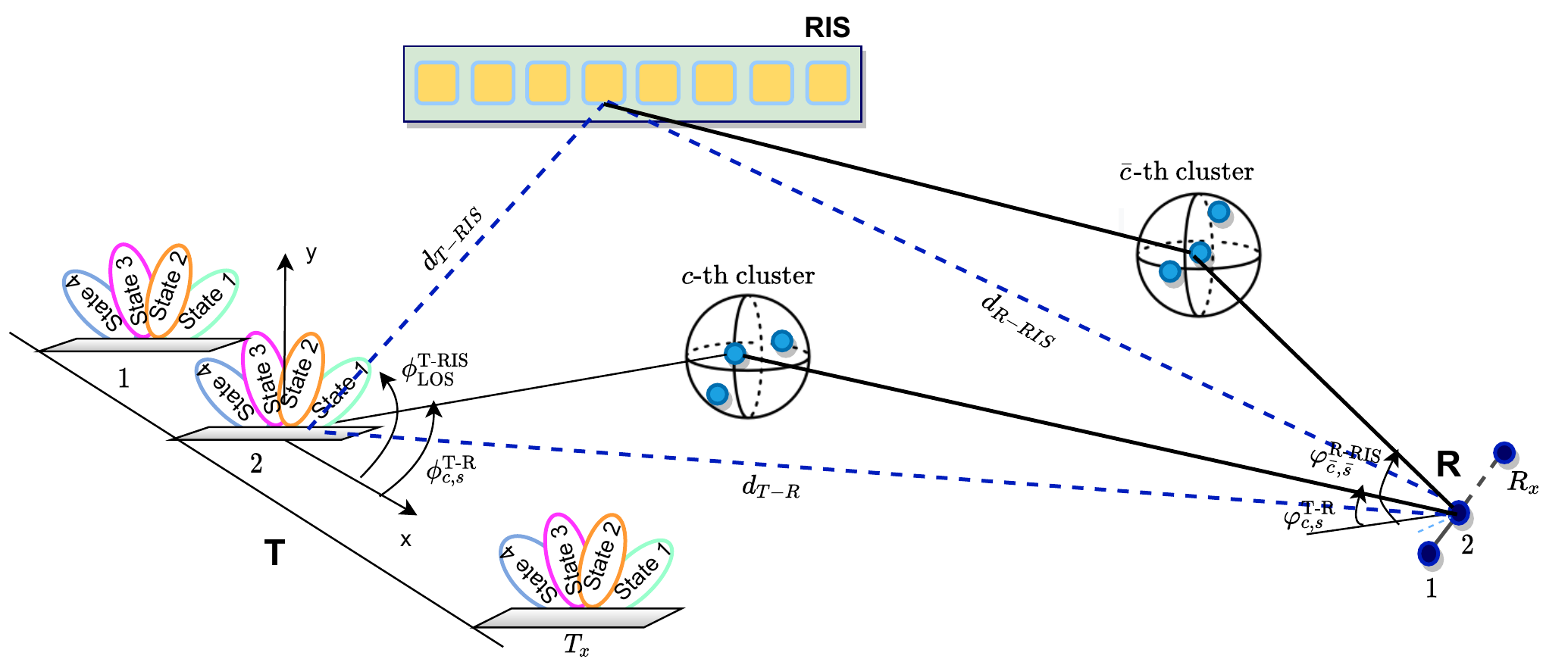}
		\vspace{-0.2cm}
		\caption{RIS-aided MIMO-MBM Systems}
		\label{p3}
		\vspace{-0.6cm}
	\end{figure}

	{Since the RIS technology offers cost-effective solutions for  future communication systems, there is a growing body of literature on novel RIS-aided  transmission system designs \cite{basar2019wireless}.} 
	In that sense, in order to enhance the signal quality of the  aforementioned MIMO-MBM schemes,  an RIS is integrated between  T and R.  
	In this new RIS-aided MBM system, to ensure  consistency with the above  2D IMT-Advanced channel models,  a ULA-based RIS  with $d_r=\lambda/2$ spaced $N$ passive reflecting elements is placed  between T and R of the $R_x\times T_x$ MIMO-MBM system as illustrated in  Fig. \ref{p3}.   Furthermore,  in this system, in order to get a significant gain from the RIS, { we assume   the RIS being close enough to T to ensure pure LOS links}. On the other hand,  R is assumed to be sufficiently far from the RIS and T to allows multi-path propagation in T-R and R-RIS links. 
	
	Obviously,  the multi-path propagation environment of the T-R link corresponds to the MIMO-MBM channel described in the previous subsection. Therefore, the channel  matrix  characterizing   the  T-R link becomes $\mathbf{F}\in\mathbb{C}^{R_x\times T_x}$  as given in (\ref{mimo})-(\ref{2mimo}). 
	
	Likewise, since we assume pure LOS links   in T-RIS link, considering IMT-Advanced models, the LOS channel  of T-RIS link is obtained as   
	\begin{align}
		&\hspace{-1cm}	\mathbf{H}=	I_{\text{LOS}}(d_{\text{T-RIS}})\sqrt{P_L(d_{\text{T-RIS}})G_e} e^{j\eta}\nonumber\\
		&	\hspace{2cm}\times\mathbf{a}(\phi_{\text{\text{LOS}}}^{{\text{T-RIS}}})\mathbf{a}^{\mathrm{H}}(\varphi_{\text{\text{LOS}}}^{\text{T-RIS}})
		\label{new}
	\end{align}
	where $d_{\text{T-RIS}}$ is the LOS distance of the T-RIS path, and $G_e$ symbolizes the gain of the RIS elements, which  is assumed to be $G_e=\pi$ {\cite{basar2020indoor}}. In addition,  $P_L({d_{\text{T-RIS}}})$ is the attenuation between T-RIS link that can be calculated from  (\ref{pl}), while $\mathbf{a}(\phi_{\text{{LOS}}}^{{\text{T-RIS}}})$ and $\mathbf{a}(\varphi_{\text{\text{LOS}}}^{\text{T-RIS}})$ respectively are the array response vectors of the T and  RIS  in LOS direction. Thus, $\mathbf{a}(\phi_{\text{{LOS}}}^{{\text{T-RIS}}})$ is calculated as in (\ref{1mimo}), $\mathbf{a}(\varphi_{\text{\text{LOS}}}^{\text{T-RIS}})$ is obtained as follows
		\begin{equation}
		\mathbf{a}(\varphi_{\text{LOS}}^{\text{T-RIS}})=\begin{bmatrix}
			1&e^{jkd_r\sin(\varphi_{\text{LOS}}^{\text{T-RIS}})}&\cdots&e^{jkd_r(N-1)\sin(\varphi_{\text{LOS}}^{\text{T-RIS}})}
		\end{bmatrix}^T.
		\label{ris}
	\end{equation}

	On the other hand, for the R-RIS link,  we consider    a multi-path propagation environment. Let  $\bar{C}$ and $\bar{S}$ respectively symbolize the number of clusters and scatterers within each cluster in R-RIS link. Then, the  R-RIS channel matrix is constructed as 
	\begin{align}
		&	\hspace{-0.4cm}	\mathbf{G}=\sum_{c=1}^{\bar{C}}\sum_{s=1}^{\bar{S}}\alpha_{\bar{c},\bar{s}}{\sqrt{G_e{P_L({d_{\text{R-RIS}}})}}}\nonumber\\&\hspace{1.5cm}\times\mathbf{a}(\phi_{\bar{c},\bar{s}}^{{\text{R-RIS}}})\mathbf{a}^H(\varphi_{\bar{c},\bar{s}}^{\text{R-RIS}})+\mathbf{G}_{\text{LOS}}
	\end{align}
	where $d_{\text{R-RIS}}$ is the distance from the RIS to R, {$\mathbf{G}_{\text{LOS}}$} is the LOS component which is separately { obtained from multi-path propagation}, and    $\mathbf{a}(\phi_{\bar{c},\bar{s}}^{{\text{R-RIS}}})$ and $\mathbf{a}(\varphi_{\bar{c},\bar{s}}^{\text{R-RIS}})$ respectively are the array response vectors at $\phi_{\bar{c},\bar{s}}^{\text{R-RIS}}$ and  $\varphi_{\bar{c},\bar{s}}^{\text{R-RIS}}$ directions. 	
	
	After all, the composite channel of the  RIS-aided MBM system is constructed by combinations of the direct and the reflected paths  as \mbox{$\mathbf{C}\in\mathbb{C}^{R_x\times T_x}=\mathbf{G}\mathbf{\Phi H}+\mathbf{F}$}, where $\mathbf{\Phi}\in\mathbb{C}^{N\times N}$ is the diagonal reflection matrix of the RIS. It is worth stating that diagonal reflecting matrix of  $\mathbf{\Phi}$  is constructed considering \cite{basar2019transmission} for  SISO-MBM and \cite{yigit2020low} for MIMO-MBM systems.

	{Consequently}, the received signal  of the RIS-aided MIMO-MBM system $\mathbf{y}\in\mathbb{C}^{R_x\times 1}$ becomes
	\begin{equation}
		\mathbf{y}=\mathbf{Cs}+ \mathbf{n}
		\label{received}
	\end{equation}
	for $\mathbf{s}\in\mathbb{C}^{T_x\times 1}=\mathbf{1}$ being the all-one transmission signal vector whose each element corresponds to an unmodulated carrier signal, while $\mathbf{n}$ is additive white Gaussian noise (AWGN) vector composing i.i.d. random variables, where each  follows   $\mathcal{CN}(0,N_0)$ distribution.
	\vspace{-0.1cm}
	\section{Correlation and Capacity Analysis}	
		\vspace{-0.1cm}
	This section provides  correlation and capacity analyses of the  MBM systems.
	\subsection{ Correlation Analysis}
	\vspace{-0.1cm}
	Although  channel correlation is a key factor determining robustness and efficiency of the MBM systems, there has been little discussion about the  realistic correlation models for MBM \cite{basar2019media}.  In this study, although the same propagation environment is considered, the pattern diversity of the RA in Fig. \ref{figure2} \cite{jin2018simple} which exploits the RF characteristics of the environment, allows signals from different radiation patterns to follow different paths in arriving the receiver.  
	
	In this subsection, the  channel correlation caused by  different radiation patterns of the RA \cite{jin2018simple} is computed through Monte Carlo simulations. 	For this reason, the SISO-MBM scheme  whose   transmit channel   coefficient $h_p$ corresponding to the $p$th RA pattern  given in (\ref{siso}), is considered.  For $U$ being number of total sub-rays, the cross channel correlation  corresponding to $p$ and $q$ modes of the RA can be calculated from  
	\begin{align}
		&{{\bf{Q}}_{p,q}} = \mathbb{E}\left\{ {{h_p}h_q^{*}} \right\}\\
		& \hspace{0.8cm}= \mathbb{E}\left\{ \sum\limits_{u = 1}^U {\left\{ {{P_L}(d_{\text{T-R}})\sqrt {{G_p}(\phi _u^{\text{T-R}}){G_q}(\phi _u^{\text{T-R}})} } \right\}}  \right\}
	\end{align}
	where $p,q\in\left\lbrace1,2,3,4 \right\rbrace $, and { $\mathbb{E}\left\lbrace \cdot\right\rbrace $ is the expectation.}
	
	Above all, considering  IMT-Advanced NLOS  InH scenario, the cross correlation matrix of the SISO-MBM scheme is calculated via  extensive computer simulations    as follows
	\begin{equation}
		{{\bf{Q}}} = \begin{bmatrix}
			1.0000  &  0.7441  &  0.5667 &   0.6573\\
			0.7441 &   1.0000  &  0.7390  &  0.6645\\
			0.5667  &  0.7390  &  1.0000  &  0.9696\\
			0.6573  &  0.6645  &  0.9696  &  1.0000\\
		\end{bmatrix}. 
	\label{R}
	\end{equation}   
	The resulting $\mathbf{Q}$ correlation matrix shows  a manageable correlation between the channels generated by   the radiation patterns of the RA \cite{jin2018simple}, indicating the compatibility of the reference  RA \cite{jin2018simple} for the MBM transmission.
	\subsection{Capacity Analysis}
	The recent studies indicate that as in the classical MIMO transmission schemes,   the channel  capacity of IM-based systems is not dependent on the transmit signal constellations  \cite{mesleh2018space, narasimhan2015capacity}. Moreover, in \cite{narasimhan2015capacity},  the capacity of the   classical MIMO systems is calculated
	as an upper bound for the capacity of the  IM-based   MIMO systems.
	Therefore, considering \cite{narasimhan2015capacity}, the capacity of the RIS-aided MIMO-MBM systems   can be upper bounded using  (\ref{received}) as
	\begin{equation}
		R_c=\mathbb{E}\big\{\log_2\big[\det\big(\mathbf{I}+\frac{P_t}{N_0}\mathbf{CC}^{\mathrm{H}}\big)\big]\big\}\hspace{0.5cm}[\text{bits/s/Hz}]
	\end{equation}
	where $P_t$ is the transmission power and $\mathbf{I}$ is an identity matrix with $R_x$ dimensions. 
		\begin{figure}[t]
		\centering
		\includegraphics[width=0.97\linewidth]{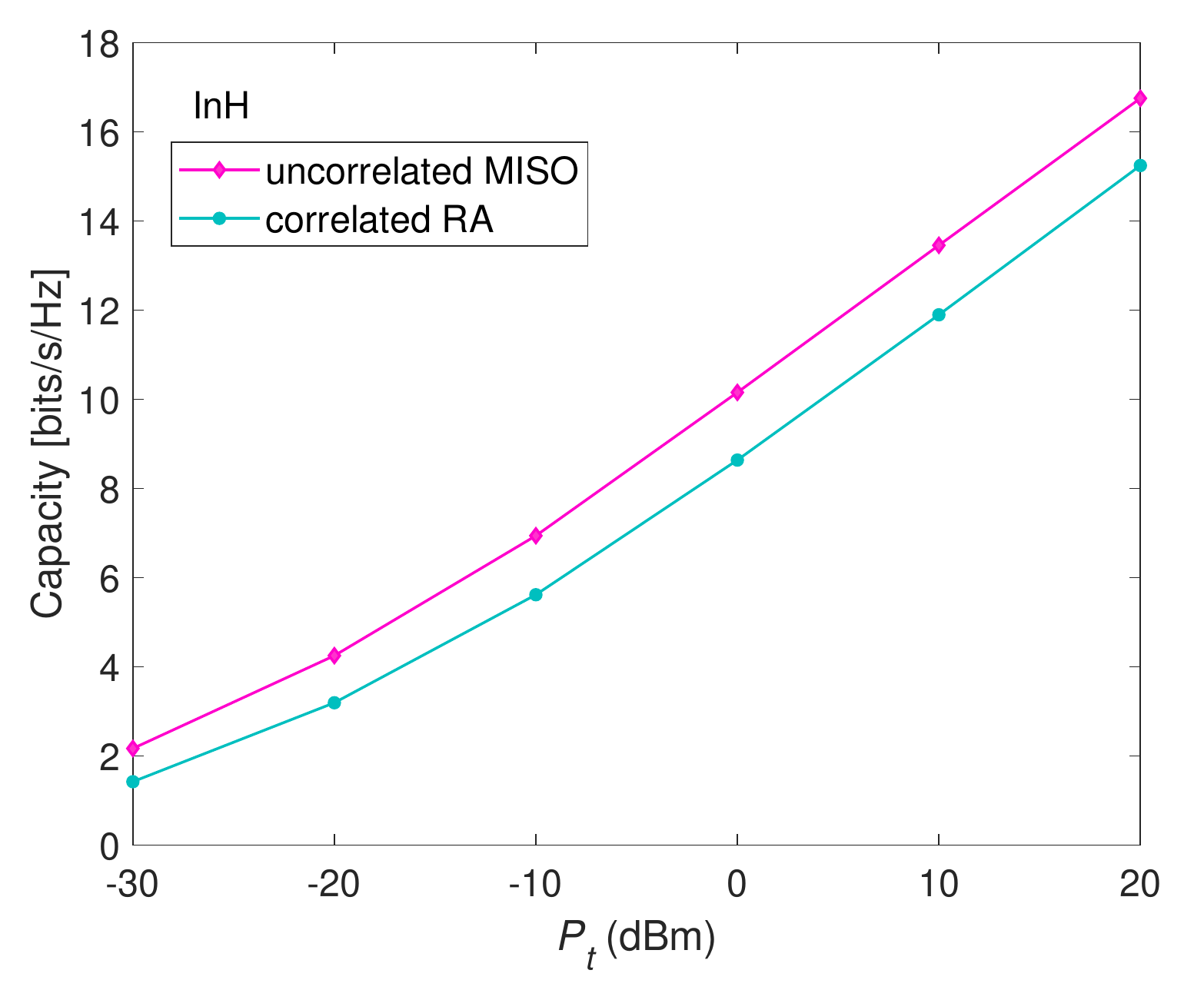}
		\caption{Channel capacity comparison of the correlated $4$-state RA \cite{jin2018simple} and uncorrelated classical MISO systems. }
		\label{uncor}
			\vspace{-0.6cm}
	\end{figure}
\enlargethispage{-1\baselineskip}
	\section{Simulation Results}
	In this section, to test  our new channel simulator, we present   a number of numerical results on the capacity of  various MBM-based systems at sub-$6$ GHz band. In all simulations,  the noise power is considered to be $-100$ dBm. Also, it should be noted that for InH scenarios,  the 2D coordinates $(x,y)$  of T, R and RIS are given as  ($30$,$10$), ($55$,$35$) and   ($33$,$13$), in meters (m), respectively. Additionally, for UMi and UMa scenarios the coordinates of T, R and  RIS are assumed to be ($45$,$15$), ($125$,$130$) and   ($48$,$18$) m.  
	
	In  Fig. \ref{uncor}, considering IMT-Advanced InH scenario, the  capacity of the RA-based system \cite{jin2018simple}  that generates  low-correlated channel variations is compared with uncorrelated classical  MISO transmission schemes that employs four transmit and a single received omni-directional antennas in transmission. The  results indicate that the performance of the  four-state RA \cite{jin2018simple} is slightly behind the uncorrelated multi-antenna system.  This comparison is significant in understanding the effect of low channel correlations on the performance of MBM systems.
	
\begin{figure*}[t]
	\centering
	\includegraphics[width=0.95\linewidth]{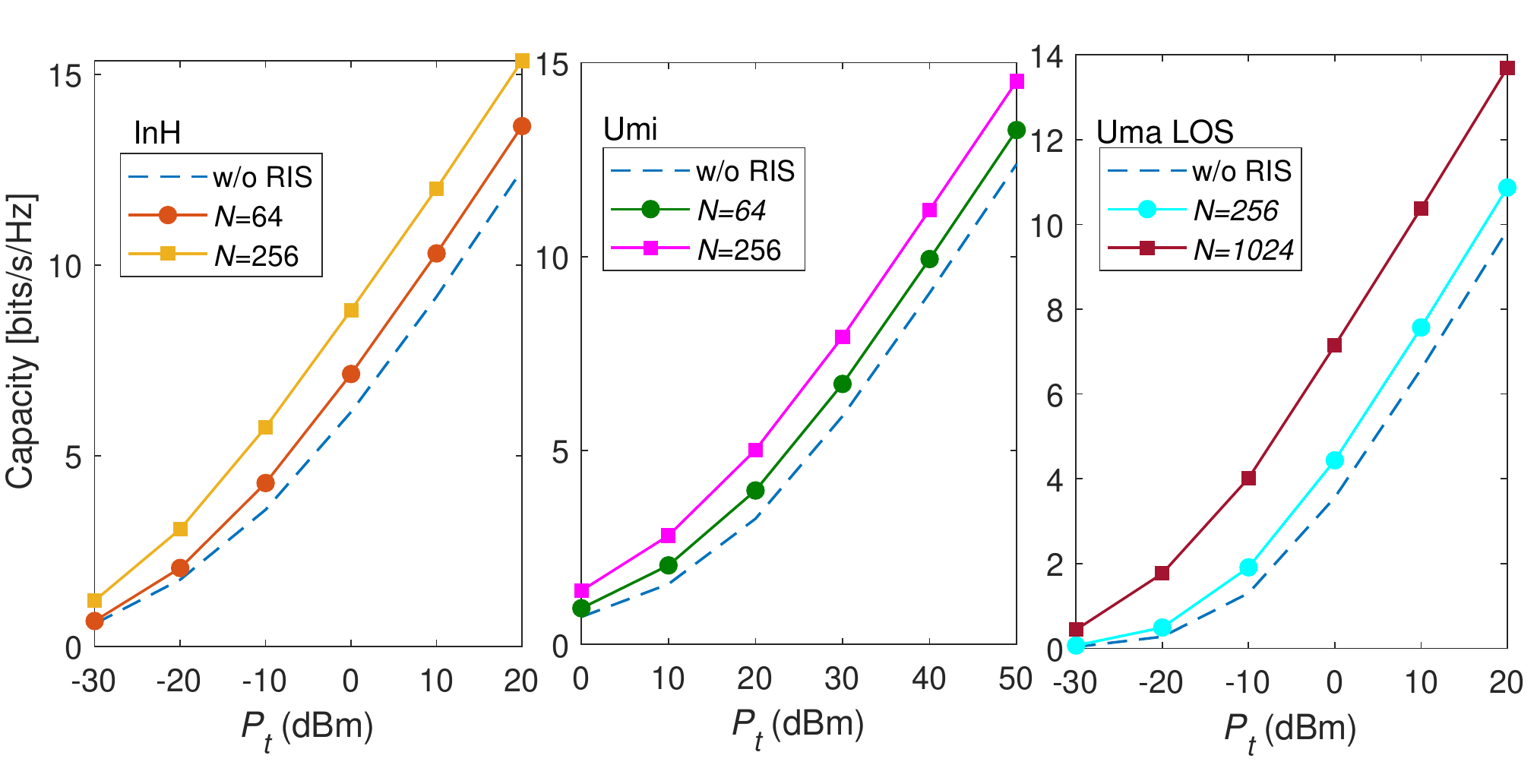}
		\vspace{-0.4cm}
	\caption{Capacity of the RIS-aided MBM systems under a) InH b) UMi c) UMa LOS scenarios for varying $N$.  }
	\label{3lu}
\end{figure*}

	In Fig. \ref{3lu}, the capacity of the  SISO-MBM with/without assistance of the RIS  is investigated  in different environments. It is clear from the results that since outdoor environments (UMi, UMa) experience higher attenuation in the direct T-R link, they require higher number of  reflecting elements in RIS to tackle  this performance degradation. 
	
\begin{figure}[t]
	\centering
	\includegraphics[width=0.95\linewidth]{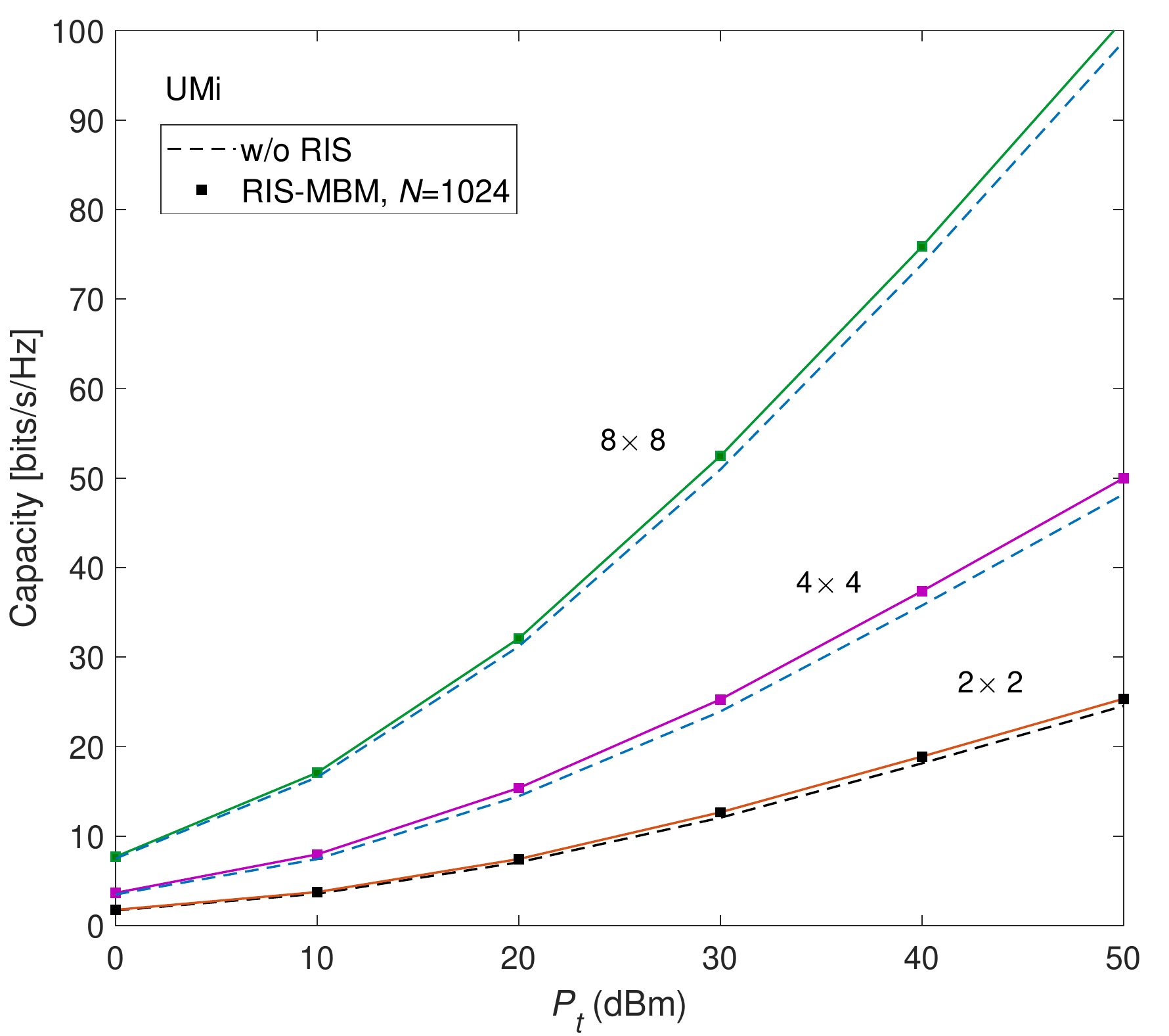}
	\caption{Capacity of the RIS aided MIMO-MBM systems for IMT-Advanced UMi scenario }
	\label{key}
		\vspace{-0.3cm}
\end{figure}

	In Fig. \ref{key}, the  capacity results of the MIMO-MBM systems in UMi environment are demonstrated. It can be deduced from the results that as MIMO configuration enlarged, a linear relation between  the path attenuation and capacity of the MIMO-MBM systems is observed. Therefore, the effect of the RIS on the system performance is hardly seen. 
\enlargethispage{-1\baselineskip}	
	\section{Conclusion}
	In this paper, a 2D, open-source and  easy-to-use physical channel simulator, \textit{SimMBM}, enabling  users to generate communication channels  of the various MBM implementations at sub-$6$ GHz frequency spectrum, has been presented. The proposed channel simulator offers flexibility in performing different MBM systems for varying localizations, scenarios and RA adaptations and reveals the potential of emerging MBM systems in realistic conditions. 
	\vspace*{-0.3cm}




	
	
	%
	\bibliographystyle{IEEEtran}
	\bibliography{refer}

\end{document}